\documentclass[11pt]{article}
   
\usepackage{graphicx,floatflt,amssymb}    
\usepackage{epsfig}    
\usepackage{graphics}    
\usepackage{psfrag}    
\newcommand{\ba}{\begin{array}}     
\newcommand{\ea}{\end{array}}     
\newcommand{\bd}{\begin{displaymath}}     
\newcommand{\ed}{\end{displaymath}}     
\newcommand{\be}{\begin{equation}}     
\newcommand{\ee}{\end{equation}}     
\newcommand{\bea}{\begin{eqnarray}}     
\newcommand{\eea}{\end{eqnarray}}

\def\ltap{\raisebox{-.4ex}{\rlap{$\sim$}} \raisebox{.4ex}{$<$}}     
\def\gtap{\raisebox{-.4ex}{\rlap{$\sim$}} \raisebox{.4ex}{$>$}}

\newcommand{\sun}{\Delta m^2_{\rm sol}}     
\newcommand{\atm}{\Delta m^2_{\rm atm}}     
     
\newcommand{\nonu}{(0\nu\beta\beta)}     
\newcommand{\meff}{|m_{ee}|}     
\newcommand{\angsun}{\sin^2 2\theta_{\rm sol}}     
\newcommand{\angchz}{\sin \theta_{13}}     
     
\begin{document}     
\vspace*{-0.5in}     
\renewcommand{\thefootnote}{\fnsymbol{footnote}}     
\begin{flushright}     
LPT Orsay/03-13 \\     
SINP/TNP/03-10\\     
\texttt{hep-ph/0304159}      
\end{flushright}     
\vskip 5pt     
\begin{center}     
{\Large {\bf Impact of CP phases on neutrinoless double beta decay}}     
\vskip 25pt     
{\bf Asmaa Abada $^{1}$}     
~and~       
{\bf Gautam Bhattacharyya $^{2}$}     
\vskip 10pt       
$^{1)}${\it Laboratoire de Physique Th\'eorique,      
Universit\'e de Paris XI, B\^atiment 210, \\ 91405 Orsay Cedex,     
France} \\     
$^{2)}${\it Saha Institute of Nuclear Physics, 1/AF Bidhan Nagar,      
Kolkata 700064, India}     
\vskip 20pt     
{\bf Abstract}     
\end{center}     
     
\begin{quotation}     
  {\noindent\small We highlight in a model independent way the
dependence of the effective Majorana mass parameter, relevant for
neutrinoless double beta decay, on the CP phases of the PMNS matrix,
using the most recent neutrino data including the cosmological WMAP
measurement. We perform our analysis with three active neutrino
flavours in the context of three kinds of mass spectra:
quasi-degenerate, normal hierarchical and inverted hierarchical. If a
neutrinoless double beta decay experiment records a positive signal,
then assuming that Majorana masses of light neutrinos are responsible
for it, we show how it might be possible to discriminate between the
three kinds of spectra.

\vskip 5pt     
\noindent     
PACS number(s):~14.60.Pq, 23.40.-s.      
}     
     
\end{quotation}     
     
\vskip 15pt       
     
\setcounter{footnote}{0}     
\renewcommand{\thefootnote}{\arabic{footnote}}

The importance of looking for neutrinoless double beta decay $\nonu$
lies in the fact that, if observed, it would establish a violation of
the total lepton number, which is otherwise a conserved quantum number
in the standard model.  Any nonvanishing amplitude for this decay may
be inferred as a signal for an effective Majorana mass of the electron
neutrino. This way it is sensitive to some kind of an absolute mass of
the neutrino, contrary to the oscillation experiments, which can fix
only the neutrino mass squared differences.  An evidence for this
decay has recently been claimed on the basis of results from
Heidelberg-Moscow experiments \cite{db_klap}.  This claim has been
criticized by authors in \cite{criticism}, which has subsequently been
followed by a reply to the criticism made \cite{klap_reply}.  In any
case, the currently running NEMO3 experiment \cite{nemo3} and future
\cite{future_db} (Majorana, EXO, CUORICINO, CUORE, GENIUS) $\nonu$
experiments would either confirm this evidence or would put a stronger
bound on the amplitude of this decay. The rate of $\nonu$ is
proportional to the square of the ($ee$)-element of the neutrino mass
matrix, often called the effective mass parameter $m_{ee}$. This
parameter depends on the absolute neutrino masses, the solar and CHOOZ
mixing angles, and two CP phases. A detailed discussion of the
dependence of $m_{ee}$ on different parameters may be found, {\em
e.g.}, in \cite{kps,petcov02}.
   
{\em The purpose of this short note} is to highlight in a model
independent way the dependence of $m_{ee}$ on the CP phases,
using the most recent oscillation data on mass square splittings and
mixing angles \cite{solar_expt,solar,atm_expt,atm,chooz,pv}, as well
as the recent cosmological bound from WMAP on the sum of all neutrino
masses \cite{wmap} in conjunction with data from 2dF galaxy redshift
survey (2dFGRS) \cite{2df}.  We base our analysis on the three
possible kinds of mass spectra: quasi-degenerate, hierarchical and
inverted hierarchical, in the context of three neutrino generations.
The $\nonu$ experiment in a sense serves to distinguish between the
spectra: due to the present sensitivity, its observation in the
ongoing $\nonu$ experiment, as it would turn out, would only establish
a nearly degenerate mass spectrum.

We stress at this point that even though the $\nonu$ amplitude does  
depend on the CP phases, this decay does not correspond to a  
manifest CP-violating phenomenon. The rate of this decay is indeed  
affected by the phases. But the effect is CP-even, {\em i.e.}, the  
rate of this decay in a nucleus will be the same, in principle, to  
that in an antinucleus. The CP-odd effect that these Majorana phases  
might cause have been studied in the context of neutrino  
$\leftrightarrow$ antineutrino oscillation, rare leptonic decays of  
the $K$ and $B$ mesons, and leptogenesis (for a recent discussion on  
this issue, see \cite{gkm}).

Let us now set up our notations in a scenario with three active     
neutrino flavours. In other words, we keep the LSND results  
\cite{lsnd}     
out of our consideration.\footnote{Indeed,    
we know now that miniBoone \cite{miniboone} will either     
confirm or rule out the LSND signal earliest by 2006,   
see \cite{Green}.}     
We recall that observation of neutrino    
oscillation    
implies mixing between the flavours due to the fact that the flavour    
basis is not parallel to the mass basis.     
The flavour basis is written as $\nu_{\ell L}$     
where $\ell = e, \mu, \tau$, and the mass basis is expressed as     
$\nu_{iL}$ where $i = 1, 2,     
3$ ($L$ stands for left-handed). 
The two bases are related to each other by      
\be \nu_{\ell L} =     
\sum_{i=1}^3 U_{\ell i} \nu_{iL},        
\ee      
where the unitary matrix $U$ is called the     
Pontecorvo-Maki-Nakagawa-Sakata (PMNS) matrix \cite{pmns}. A useful     
parametrization of $U$ is given by \cite{C-K}     
\bea\label{mix}     
U =  \left(     
 \begin{array}{ccc}     
 c_{12}\ c_{13}&s_{12}\ c_{13}& s_{13} e^{-i \delta}\\     
 -s_{12}\ c_{23}-c_{12}\ s_{23}\ s_{13}e^{i \delta}     
 &c_{12}\ c_{23}-s_{12}\ s_{23}\     
 s_{13}e^{i \delta}&s_{23}\ c_{13}\\     
 s_{12}\ s_{23}-c_{12}\ c_{23}\ s_{13}e^{i \delta}     
 &-c_{12}\ s_{23}-s_{12}\ c_{23}\ s_{13}e^{i \delta}&     
 c_{23}\ c_{13}     
 \end{array}     
 \right)\  {\mathrm{diag}}\left\{e^{i \alpha_1},e^{i \alpha_2} ,    
1\right\},     
 \eea     
where $c_{ij}=\cos (\theta_{ij})$ and $s_{ij}=\sin (\theta_{ij})$,     
$\delta$ is the Dirac CP phase and $\alpha_{1,2}$ are the Majorana     
phases.

If the  $\nonu$ amplitude is indeed generated by a $(V-A)$ weak charged  
current interaction via Majorana neutrino exchange, and if the masses  
of those neutrinos are less than a typical Fermi momentum ($\sim 100$  
MeV) of the nucleons inside a nucleus, then the $\nonu$ amplitude is  
proportional to the effective mass $m_{ee}$ defined as \cite{bp_rev}  
\be  
\label{master}     
\meff = \left|U_{ei}^2 m_i\right| = \left|m_1 |U_{e1}|^2 + m_2
|U_{e2}|^2 e^{2i\alpha_{_{M}}} + m_3 |U_{e3}|^2 e^{2i\alpha_{_{MD}}}
\right|, \ee where $\alpha_{_{M}} = (\alpha_2 - \alpha_1)$ is a pure
Majorana type and $\alpha_{_{MD}} = -(\delta + \alpha_1)$ is a mixture
of the Majorana and Dirac type CP phases. Without any loss
of generality, we can take the mass eigenvalues $(m_1, m_2, m_3)$ to
be positive. The effective mass parameter depends on the solar angle
$\theta_{12}$, the CHOOZ angle $\theta_{13}$, the masses $m_i$ and the
CP phases. The solar angle measurement has become increasingly precise
particularly after the SNO results came out. As regards $\theta_{13}$,
there exists only an upper limit from the CHOOZ \cite{chooz} and Palo
Verde \cite{pv} neutrino disappearance reactor 
experiments.  The latter angle links
the solar and the atmospheric sectors in the PMNS matrix.  This angle
is also important in the context of future CP violation measurements
in the long baseline experiments. For an observable impact of CP
violation $\theta_{13}$ should not be smaller than $0.2^\circ$ (the
other necessity is a large solar angle which has already been
established anyway).  More specifically, the future first generation
superbeams JHF-SK \cite{jhf} and NuMI \cite{numi} long baseline
experiments (JHF-SK to start taking data in 2007) along with possible
large reactor experiments will measure $\sin^2 \theta_{13}$ to a few
$10^{-3}$ level \cite{huber} and, if luck permits, will also determine
some CP asymmetries.  Now we turn our attention to the CP phases.  As
yet, these phases are completely unknown.  Only the $\nonu$ amplitude
offers a unique and direct probe to them.  These phases take an active
r\^{o}le in determining the size of the $\nonu$ amplitude, and the
possibility of a likely signal for this decay in the current and
foreseeable experiments hangs crucially on the amount of destructive
interferences created by these phases.

We now briefly summarize the experimental data which concern the    
effective mass calculation related to neutrinoless double beta decay.
\begin{itemize}      
\item The post-KamLAND analysis \cite{solar} constrain      
the solar angle,  $\theta_{\rm {sol}}$ or $\theta_{12}$, as 
(b.f. means best fit)     
\be      
\label{solarang}     
0.70~\ltap~\angsun~\ltap~0.96 ~~(95\%~{\rm CL});     
~~~\angsun~({\rm b.f.}) = 0.82.     
\ee     
     
\item The CHOOZ experiment \cite{chooz} constrains the $\theta_{13}$   
angle as      
\be      
\label{chooz}     
\angchz~\ltap~0.22 ~~(95\%~{\rm CL})\ ,      
\ee     
 and a global analysis by Fogli {\em et al.} \cite{Fogli} led to   
  $|U_{e3}|^2 < 5.0\cdot 10^{-2}~~(99.7$\%~{CL}).    
\item The solar \cite{solar} and atmospheric \cite{atm} squared mass     
differences are constrained as (95\% CL)      
\bea      
\label{splitt}     
5.8 \cdot 10^{-5}~\ltap~\sun~({\rm eV}^2) ~\ltap~ 9.1 \cdot 10^{-5} \;  
~~~     
\sun~({\rm b.f.}) = 7.2 \cdot 10^{-5} ~({\rm eV}^2) \; \\        
1 \cdot 10^{-3}~\ltap~\atm~({\rm eV}^2) ~\ltap~ 5.0 \cdot 10^{-3} \;  
~~~     
\atm~({\rm b.f.}) = 2.5 \cdot 10^{-3} ~({\rm eV}^2) .         
\eea     
     
\item The WMAP result \cite{wmap} in conjunction with the 2dFGRS data     
\cite{2df} constrain the total mass of the active neutrino species 
(with     
the assumption that these neutrinos have decoupled while still being     
relativistic) as      
\be      
\label{wmap}     
\sum_i m_i ~\ltap~ 0.71 ~{\rm eV} ~~(95\% ~{\rm CL}).      
\ee      
Implicitly, the limit in Eq.~(\ref{wmap}) uses the Ly-$\alpha$ forest    
data \cite{forest}    
whose interpretation is still controversial. Excluding the    
latter, one obtains a more robust and conservative bound     
$ \sum_i m_i ~\ltap~ 1.01 ~{\rm eV}$ \cite{hannestad}.

\item The Heidelberg-Moscow claim on evidence of $\nonu$ translates    
into an effective Majorana mass \cite{db_klap,criticism,klap_reply}    
\be 0.11~\ltap~ \meff~({\rm eV}) ~\ltap~0.56 ~~(95\% ~{\rm CL}); ~~~    
\meff~({\rm b.f.}) = 0.39 ~{\rm eV}.  \ee

\item The Mainz \cite{mainz} and Troitsk \cite{troitsk} Tritium beta   
decay experiments have put the bound    
$m_{\nu_e} ~\ltap~ 2.2$ eV   
on the electron-type   
neutrino mass. The future KATRIN Tritium beta decay experiment   
\cite{katrin}, planned to be operative from 2007, has the possibility   
to probe $m_{\nu_e}$ down to 0.35 eV level.      
\end{itemize}      
     
We perform our analysis on the basis of the usual three kinds of mass 
hierarchy, and we discuss them one by one.  But, before that, we  
observe  that the WMAP limit automatically sets an {\em upper 
limit} for the effective mass parameter in neutrinoless double beta 
decay. In other words, keeping in mind Schwarz inequality, it follows 
from Eq.~(\ref{master}) that $\meff ~\ltap~ 0.71 $ eV 
(or a more conservative upper limit of 1.01 eV {\em a la} Hannestad 
\cite{hannestad}). A similar conclusion was also drawn in \cite{am}. 
 
\begin{enumerate}      
\item \underline{Quasi-degenerate}:~ The three eigenvalues are $m_1    
\simeq m_2 \simeq m_3 \equiv m_0$. The absolute scale can be made    
large enough to saturate the WMAP bound, i.e. $m_0 \simeq 0.23$ eV. In    
this case, Eq.~(\ref{master}) turns out to be \be    
\label{deg}      
\meff \simeq m_0 \left|c_{12}^2 c_{13}^2+ s_{12}^2 c_{13}^2  
e^{2i\alpha_{_M}} + s_{13}^2 e^{2i\alpha_{_{MD}}}\right|.    
\ee   
 
Since CHOOZ data constrain $s_{13}$ to be small, one would naively
throw away the third term in Eq.~(\ref{deg}), as has been the usual
practice.  But the effect of this term can be significant when there
is a cancellation between the first two terms. For $s_{13} = 0$, we
obtain 
\be 
m_0 |\cos 2\theta_{\rm sol}| ~\ltap~ \meff ~\ltap~ m_0\ .
\label{deg-lim} 
\ee 
As a matter of fact, the upper bound $m_0$ in Eq.~(\ref{deg-lim}) 
holds, thanks to the Schwartz inequality, irrespective of the value
of $s_{13}$. The r\^{o}le of the
destructive interference can be seen in Fig.~1 where we have plotted
the effective mass parameter for both $s_{13} = 0$ (left panel) and
the maximum allowed value $s_{13} = 0.22$ (right panel). We point out
here that the lowest value of $\meff/m_0$ in the plots of Fig.~1 is
not zero but $|\cos 2\theta|_{\rm sol}$.  The relative importance of
the two phases and also the impact of nonvanishing $\sin \theta_{13}$
(right panel) are apparent from Fig.~1. Comparing the left and right
panels, we infer that a non-vanishing $s_{13}$ (we put the CHOOZ upper
limit of 0.22) somewhat suppreses (by something like 10\%) 
the maximum value $m_0$ can attain 
for $\alpha_M = 0$ compared to the $s_{13} = 0$ scenario.

\item \underline{Normal Hierarchy}:~ In this case, $m_1 < m_2 \ll   
m_3$. As an illustrative example, we can take $m_1 \simeq 0$, $m_2   
\simeq \sqrt{\sun}$, and $m_3 \simeq \sqrt{\atm}$. Then one can   
effectively get rid of one of the two CP phases in Eq.~(\ref{master}),   
and can write ($\alpha \equiv \alpha_{_{MD}} - \alpha_{_{M}}$)    
\be   
\label{nh}      
\meff = \left|\sqrt{\sun}~ s_{12}^2 c_{13}^2+ \sqrt{\atm}~ s_{13}^2     
~e^{2i\alpha}\right|.     
\ee     
   
In Fig.~2, we have plotted ${\meff/\sqrt{\sun}}$ against the  
CP phase $\alpha$. We observe that even in the  
case of  maximum cancellation  
($\alpha=\pi/2$) the effective mass never vanishes (see the zoom in  
Fig.~2) and thus corresponds to 
a lower bound, which is unfortunately much  
below the present and foreseeable  experimental sensitivity.  Putting  
numbers, we obtain within the 95\% confidence level from the data 
 \be  
  \meff ~\ltap~ 0.007  
  ~{\rm eV}\ .  
  \ee  
A non-zero $m_1$ (but small enough to satisfy $m_1  
\ll \sqrt{\sun}$) can however push $\meff$ to slightly higher values.
   
\item \underline{Inverted Hierarchy}:~ In this case, $m_1 > m_2 \gg     
m_3$. One can take $m_1 \simeq m_2 \simeq \sqrt{\atm}$ and $m_3 
\simeq     
0$. Again, only one CP phase, the pure Majorana one,  
enters into the expression for $\meff$,     
given by      
\be      
\label{ih}      
\meff = \sqrt{\atm} c_{13}^2\left|c_{12}^2 + s_{12}^2 
~e^{2i\alpha_{_{M}}}\right|.   
\ee  
This case is very similar to the 
quasi-degenerate scenario, except that the overall mass scale is 
suppressed by $\sqrt{\atm}/m_0$ and that the third term in 
Eq.~(\ref{master}) is even further suppressed.  This case is 
illustrated in Fig.~3 where we have plotted ${\meff/\sqrt{\atm}}$ as a 
function of the CP phase $\alpha_{_{M}}$. The maximum cancellation 
holds for $\alpha_{_{M}}=\pi/2$, as it was   
for the quasi-degenerate 
case.  
When $\alpha_M = 0$, which does not necessarily mean that the original
Majorana phases $\alpha_1$ and $\alpha_2$ in the PMNS matrix
individually vanish,  
we obtain the maximum  
amplitude.  Again putting numbers, we obtain at 95\% CL 
from experimental data 
\be 
0.006 ~{\rm eV} ~\ltap~ \meff ~\ltap~ 0.07 ~{\rm eV}\ . 
\ee 
 
\end{enumerate}    

Thus we may observe that a measurement of $\meff$ may serve to
distinguish between the spectra.  As an example, any measurement of
$\meff$ reasonably above the maximum $\sqrt{\atm} \simeq 0.07$ eV will
conclusively rule in favour of the quasi-degenerate spectrum,
irrespective of the present uncertainty over the absolute mass upper
limit. In the future experiments, if the effective mass is found
between 0.007 eV and 0.07 eV, then the spectrum would correspond to
inverted hierarchy pattern, while an observation of $\meff$ below
0.006 eV would imply a normal hierarchical pattern. But if $\meff$ is
observed between 0.006 and 0.007 eV, then the two kinds of hierarchies
cannot be discriminated. These divisions are based on the basis of
accepting the experimentally allowed regions at 95\% CL. If, instead,
one employs 99\% CL criterion, the lower bound of $\meff$ in the case
of inverted hierarchy enters {\em more} into the zone admitted by
normal hierarchy.  Another point to note is that in future if the
KATRIN Tritium beta decay experiment confirms a large ($\gtap~0.35$
eV) absolute mass, then a measurement of $\meff$ in an ongoing
neutrinoless double beta decay experiment would provide an idea about
the CP phases.  The Heidelberg-Moscow and NEMO3 experiments have been
designed to reach a sensitivity of a few $10^{-1}$ eV. Thus a positive
signal in these experiments will only imply a degenerate spectrum.
Among the future short term projects, CUORICINO will have a
sensitivity of a few $10^{-1}$ eV, but the Majorana, EXO, and CUORE
experiments are expected to attain a sensitivity of few $10^{-2}$
eV. Therefore, we will be able to distinguish between the inverted
hierarchical and the degenerate spectra. On the other hand, if the
spectrum is normal hierarchical then we will have to wait for the long
term projects, which are expected to reach a sensitivity of few
$10^{-3}$ eV ({\em e.g.} 10t GENIUS).  We refer to ref.~\cite{future}
for a general discussion about the future direct neutrino mass
measurements.
 
A word of caution is relevant here. Nonzero Majorana masses of light
neutrinos are not necessarily the only source behind a nonvanishing
$\nonu$. Heavy Majorana neutrinos or doubly charged scalars may also
contribute to $\nonu$, where the contributions are suppressed by their
heavy masses. In fact, in the context of left-right symmetric ${\rm
SU(2)}_L \times{\rm SU(2)}_R \times {\rm U(1)}_{B-L}$ model, the
see-saw generated light Majorana neutrinos, the heavy Majorana
neutrino, the doubly charged scalar, all contribute to $\nonu$;
additionally, there is a fourth contribution arising out of light and
heavy neutrino mixing. Non-observation of $\nonu$ can therefore be
translated into lower bounds on the relevant heavy masses in the range
of a few hundred GeV to a few TeV. In supersymmetric models with
broken R-parity, the trilinear $\lambda'_{111}$ coupling or the
product couplings $\lambda'_{11j} \cdot \lambda'_{1j1}$ also drive
$\nonu$, and again stringent bounds emerge on those couplings.  The
R-parity violating couplings will have distinct collider signals.  So
before one interprets a nonzero signal of $\nonu$ as a {\em direct}
consequence of light neutrino Majorana masses, one must ensure that
all other lepton number violating contributions are comparatively
dwarfed. It should, however, be noted that regardless of whatever
mechanism is responsible for $\nonu$, once there is a lepton number
violating interaction, neutrino Majorana masses will be definitely
generated at higher loops, even if it is forbidden at tree level. Thus
a nonvanishing $\nonu$ amplitude effectively implies a nonvanishing
neutrino Majorana mass, directly or indirectly. For an illustrative
discussion on different kinds of lepton number violating processes and
their contributions to $\nonu$, see ref.~\cite{Mohapatra_Pal}.
     
In conclusion, assuming that the Majorana masses of light neutrinos
are mainly responsible for $\nonu$, the major ingredients for the
prediction of $\meff$ are the solar and atmospheric mass splittings
(for normal and inverted hierarchical cases), the absolute mass scale
(for degenerate case), the solar mixing angle, the CHOOZ angle and the
CP phases.  The ongoing oscillation experiments provide mass squared
splittings and mixing angles.  In the near future the precision of all
the oscillation parameters will be significantly enhanced, which will
sharpen the $\meff$ prediction.  Then the chances of getting a
positive signal in the $\nonu$ experiments will depend crucially on
the CP phases. It was our aim to demonstrate the r\^{o}le of these
phases in this context.  Here we have not indulged ourselves in the
discussion of theoretical uncertainties associated with $\meff$
prediction. The uncertainty in the nuclear matrix element calculations
is estimated to be roughly ${\cal{O}} (2)$ (for a recent analysis on
theoretical and experimental uncertainties associated with $\meff$,
see, {\em e.g.}, \cite{petcov02}).  Eventually, if a non-zero $\nonu$
signal is observed in experiment, then its size will give a hint on
the nature of the spectrum.  This is an advantage over the oscillation
experiments.  Additionally, such an event will give us a handle on the
magnitude of the CP phases, which might lead to CP odd effects at an
observable level \cite{gkm}. Finally, we point out that following the
WMAP results \cite{wmap} a lot of enthusiasm has been generated
towards a close scrutiny of neutrinoless double beta decay (some of
these references are contained in \cite{postwmap}).

{\bf \underline{Acknowledgments}:} We thank C. Augier and S. Jullian
from NEMO3 collaboration, and also J.-P. Leroy for a useful discussion
and for suggesting improvements of the manuscript.  GB acknowledges
hospitality of LPT, Univ. de Paris XI, Orsay, where the work has been
initiated. GB's research has been supported, in part, by the DST,
India, project number SP/S2/K-10/2001.
   

\begin{figure}[ht]     
\hspace*{-1.8cm}     
\begin{center}     
\begin{tabular}{cc}      
\psfrag{zz}{}     
\psfrag{xx}{\small${\alpha_{_{M}}}$}     
\psfrag{yy}{\small${\alpha_{_{MD}}}$}     
\psfrag{AA}{\tiny$-{\pi\over 2}$}     
\psfrag{BB}{\tiny$-{\pi\over 4}$}     
\psfrag{vv}{\tiny$0$}     
\psfrag{CC}{\tiny${\pi\over 4}$}     
\psfrag{DD}{\tiny${\pi\over 2}$}     
\psfrag{ee}{\tiny$-{\pi\over 2}$}     
\psfrag{ff}{\tiny$-{\pi\over 4}$}     
\psfrag{ww}{\tiny$0$}     
\psfrag{gg}{\tiny${\pi\over 4}$}     
\psfrag{hh}{\tiny${\pi\over 2}$}     
\includegraphics[scale=0.70]{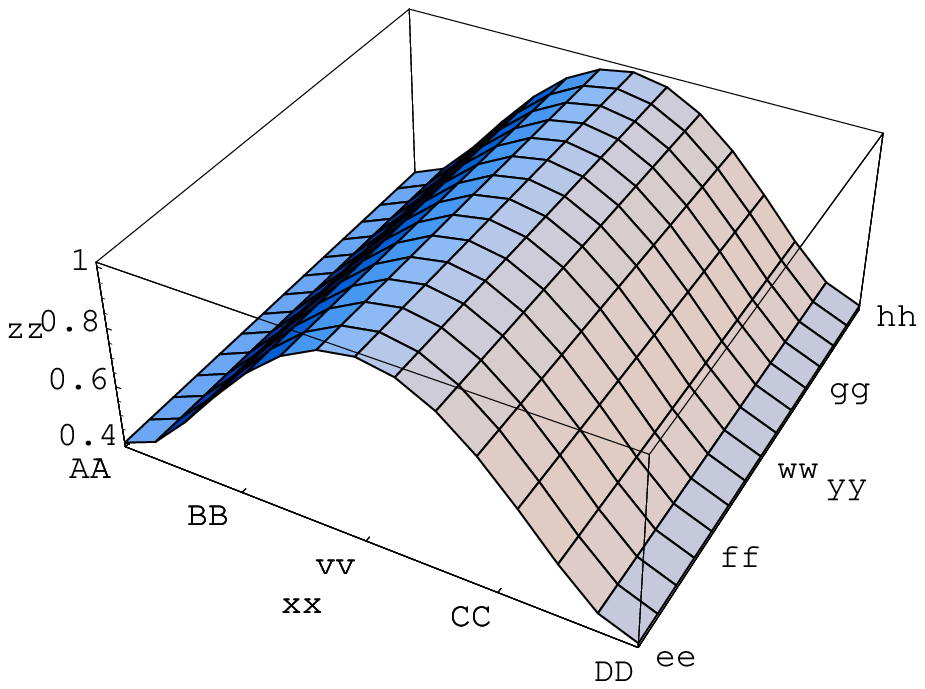} &     
\psfrag{zz}{}     
\psfrag{xx}{\small${\alpha_{_{M}}}$}     
\psfrag{yy}{\small${\alpha_{_{MD}}}$}     
\psfrag{AA}{\tiny$-{\pi\over 2}$}     
\psfrag{BB}{\tiny$-{\pi\over 4}$}     
\psfrag{vv}{\tiny$0$}     
\psfrag{CC}{\tiny${\pi\over 4}$}     
\psfrag{DD}{\tiny${\pi\over 2}$}     
\psfrag{ee}{\tiny$-{\pi\over 2}$}     
\psfrag{ff}{\tiny$-{\pi\over 4}$}     
\psfrag{ww}{\tiny$0$}     
\psfrag{gg}{\tiny${\pi\over 4}$}     
\psfrag{hh}{\tiny${\pi\over 2}$}     
\includegraphics[scale=0.70]{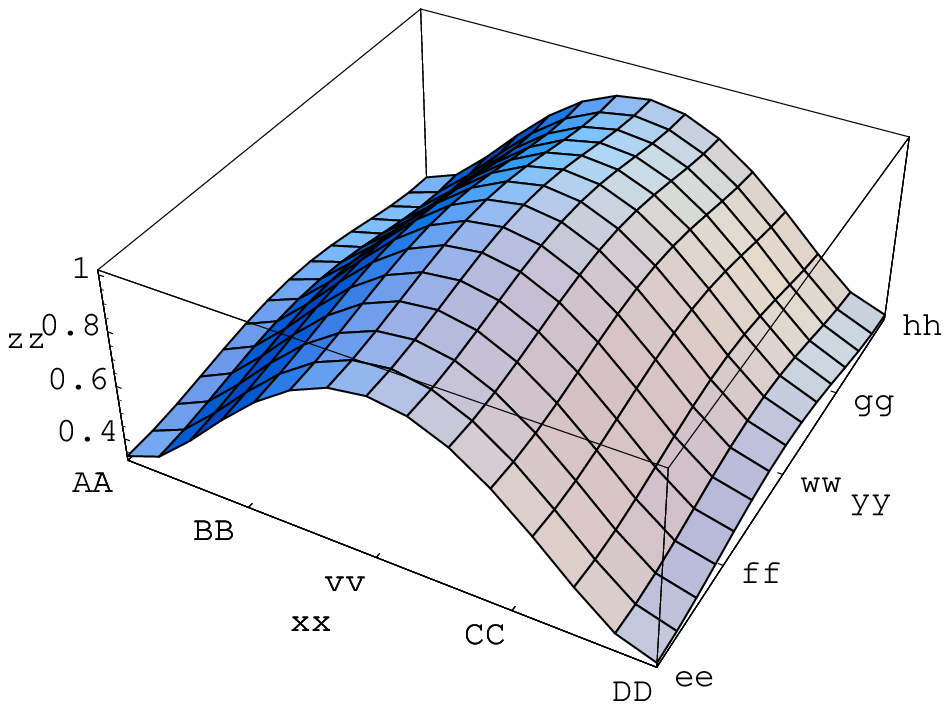}     
 \\     
 \psfrag{zz}{}     
 \psfrag{xx}{\small${\alpha_{_{M}}}$}     
\psfrag{yy}{\small${\alpha_{_{MD}}}$}     
\psfrag{AA}{\tiny$-{\pi\over 2}$}     
\psfrag{BB}{\tiny$-{\pi\over 4}$}     
\psfrag{vv}{\tiny$0$}     
\psfrag{CC}{\tiny${\pi\over 4}$}     
\psfrag{DD}{\tiny${\pi\over 2}$}     
\psfrag{ee}{\tiny$-{\pi\over 2}$}     
\psfrag{ff}{\tiny$-{\pi\over 4}$}     
\psfrag{ww}{\tiny$0$}     
\psfrag{gg}{\tiny${\pi\over 4}$}     
\psfrag{hh}{\tiny${\pi\over 2}$}     
\includegraphics[scale=0.70]{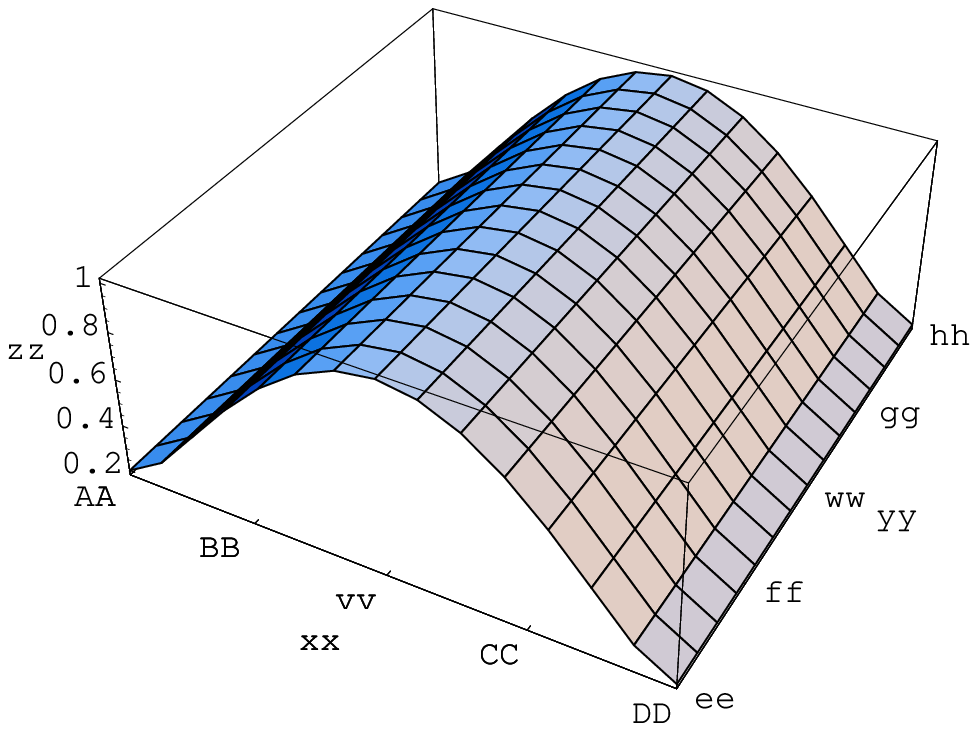}     
 &       
 \psfrag{zz}{}     
 \psfrag{xx}{\small${\alpha_{_{M}}}$}     
\psfrag{yy}{\small${\alpha_{_{MD}}}$}     
\psfrag{AA}{\tiny$-{\pi\over 2}$}     
\psfrag{BB}{\tiny$-{\pi\over 4}$}     
\psfrag{vv}{\tiny$0$}     
\psfrag{CC}{\tiny${\pi\over 4}$}     
\psfrag{DD}{\tiny${\pi\over 2}$}     
\psfrag{ee}{\tiny$-{\pi\over 2}$}     
\psfrag{ff}{\tiny$-{\pi\over 4}$}     
\psfrag{ww}{\tiny$0$}     
\psfrag{gg}{\tiny${\pi\over 4}$}     
\psfrag{hh}{\tiny${\pi\over 2}$}     
\includegraphics[scale=0.70]{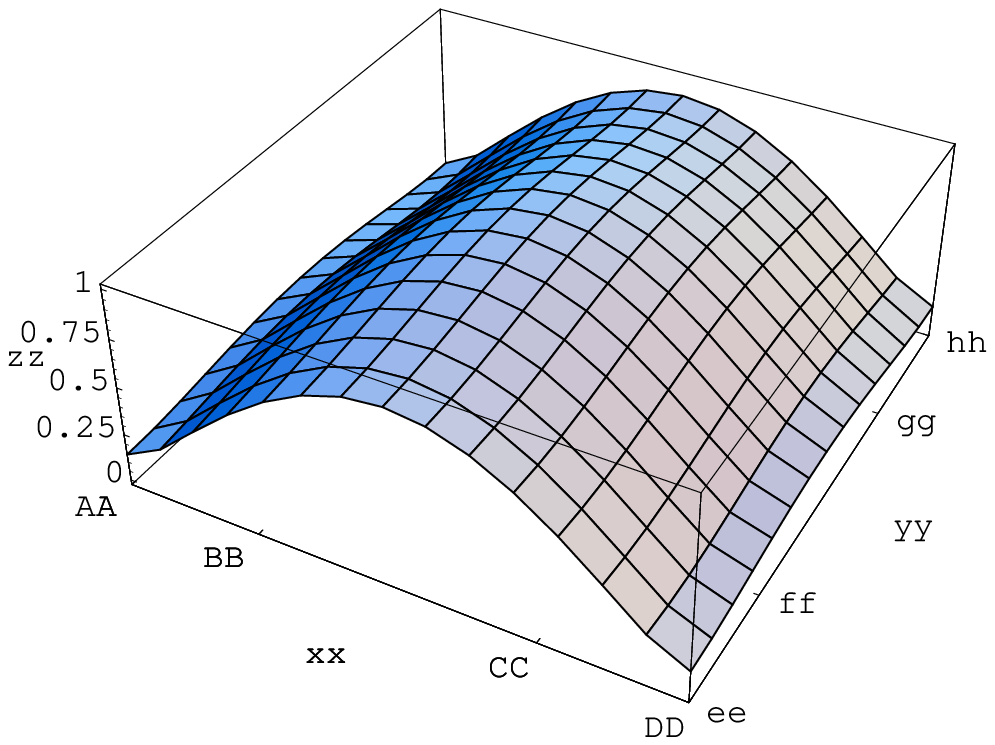}     
\\     
\psfrag{zz}{}     
\psfrag{xx}{\small${\alpha_{_{M}}}$}     
\psfrag{yy}{\small${\alpha_{_{MD}}}$}     
\psfrag{AA}{\tiny$-{\pi\over 2}$}     
\psfrag{BB}{\tiny$-{\pi\over 4}$}     
\psfrag{vv}{\tiny$0$}     
\psfrag{CC}{\tiny${\pi\over 4}$}     
\psfrag{DD}{\tiny${\pi\over 2}$}     
\psfrag{ee}{\tiny$-{\pi\over 2}$}     
\psfrag{ff}{\tiny$-{\pi\over 4}$}     
\psfrag{ww}{\tiny$0$}     
\psfrag{gg}{\tiny${\pi\over 4}$}     
\psfrag{hh}{\tiny${\pi\over 2}$}     
\includegraphics[scale=0.70]{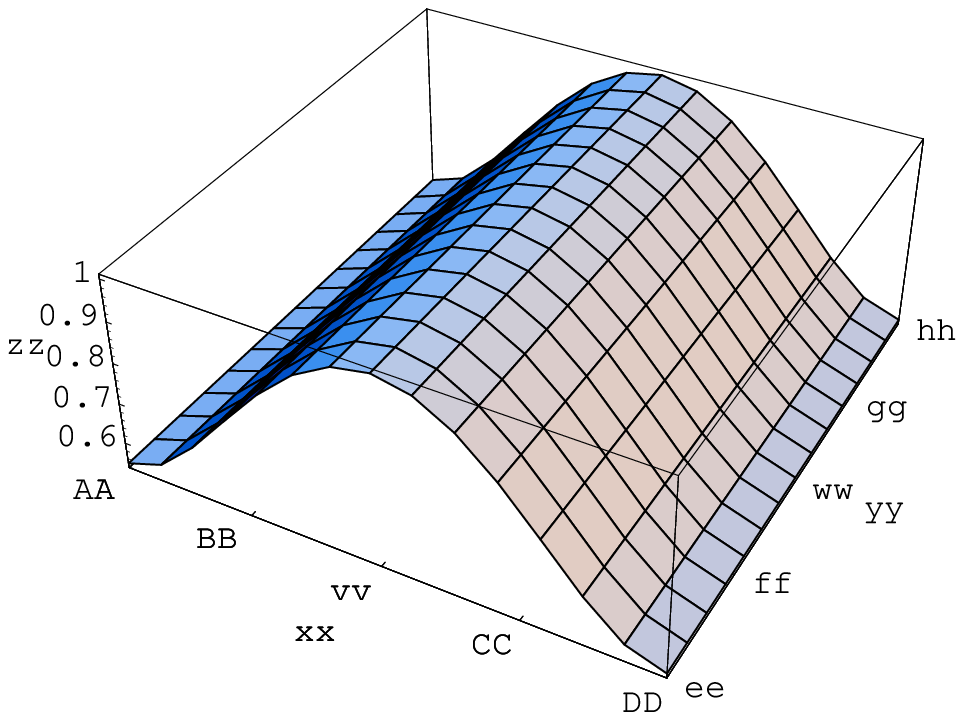}     
&\psfrag{zz}{}     
\psfrag{xx}{\small${\alpha_{_{M}}}$}     
\psfrag{yy}{\small${\alpha_{_{MD}}}$}     
\psfrag{AA}{\tiny$-{\pi\over 2}$}     
\psfrag{BB}{\tiny$-{\pi\over 4}$}     
\psfrag{vv}{\tiny$0$}     
\psfrag{CC}{\tiny${\pi\over 4}$}     
\psfrag{DD}{\tiny${\pi\over 2}$}     
\psfrag{ee}{\tiny$-{\pi\over 2}$}     
\psfrag{ff}{\tiny$-{\pi\over 4}$}     
\psfrag{ww}{\tiny$0$}     
\psfrag{gg}{\tiny${\pi\over 4}$}     
\psfrag{hh}{\tiny${\pi\over 2}$}     
\includegraphics[scale=0.70]{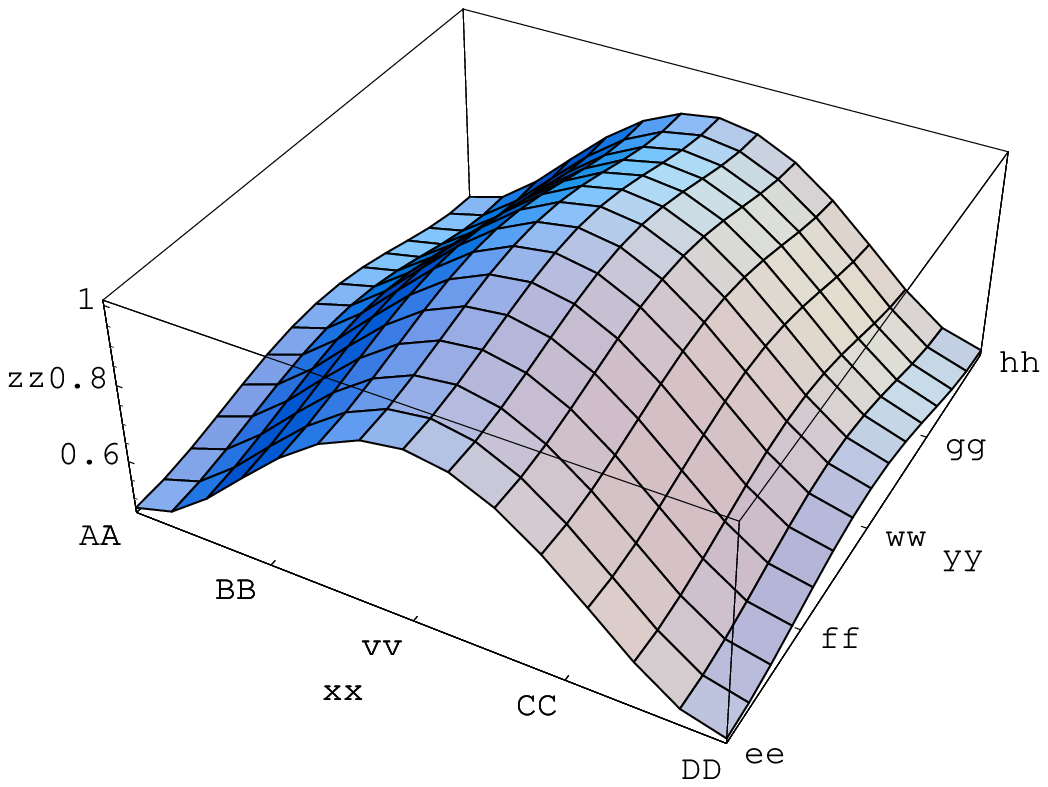}     
 \\     
\end{tabular}     
     
\end{center}     
\caption{\small The quasi-degenerate case is represented with the
minimum and maximum allowed values for the CHOOZ angle. The left panel
corresponds to $\sin \theta_{13}=0$ and the right panel to $\sin
\theta_{13}=0.22$.  The z axis represents ${\meff/m_0}$ in terms of
the two CP phases.  The lowest value of $\meff/m_0$ is not zero but
$|\cos 2\theta|_{\rm sol}$.  The first, second and third rows
correspond to $\angsun =$ 0.96 (95\% CL upper limit), 0.82 (best fit)
and 0.70 (95\% CL lower limit), respectively.}  \label{fig:degenerate}
\end{figure}
  \begin{figure}     
 \centering     
\psfrag{xx}{\small ${\alpha}$}     
\psfrag{yy}{}     
\psfrag{AA}{$0$}     
\psfrag{BB}{\tiny${\pi\over 4}$}     
\psfrag{CC}{\tiny${\pi\over 2}$}     
\psfrag{DD}{\tiny${3\pi\over 4}$}     
\psfrag{EE}{\tiny${\pi}$}     
\includegraphics[width=3.2in]{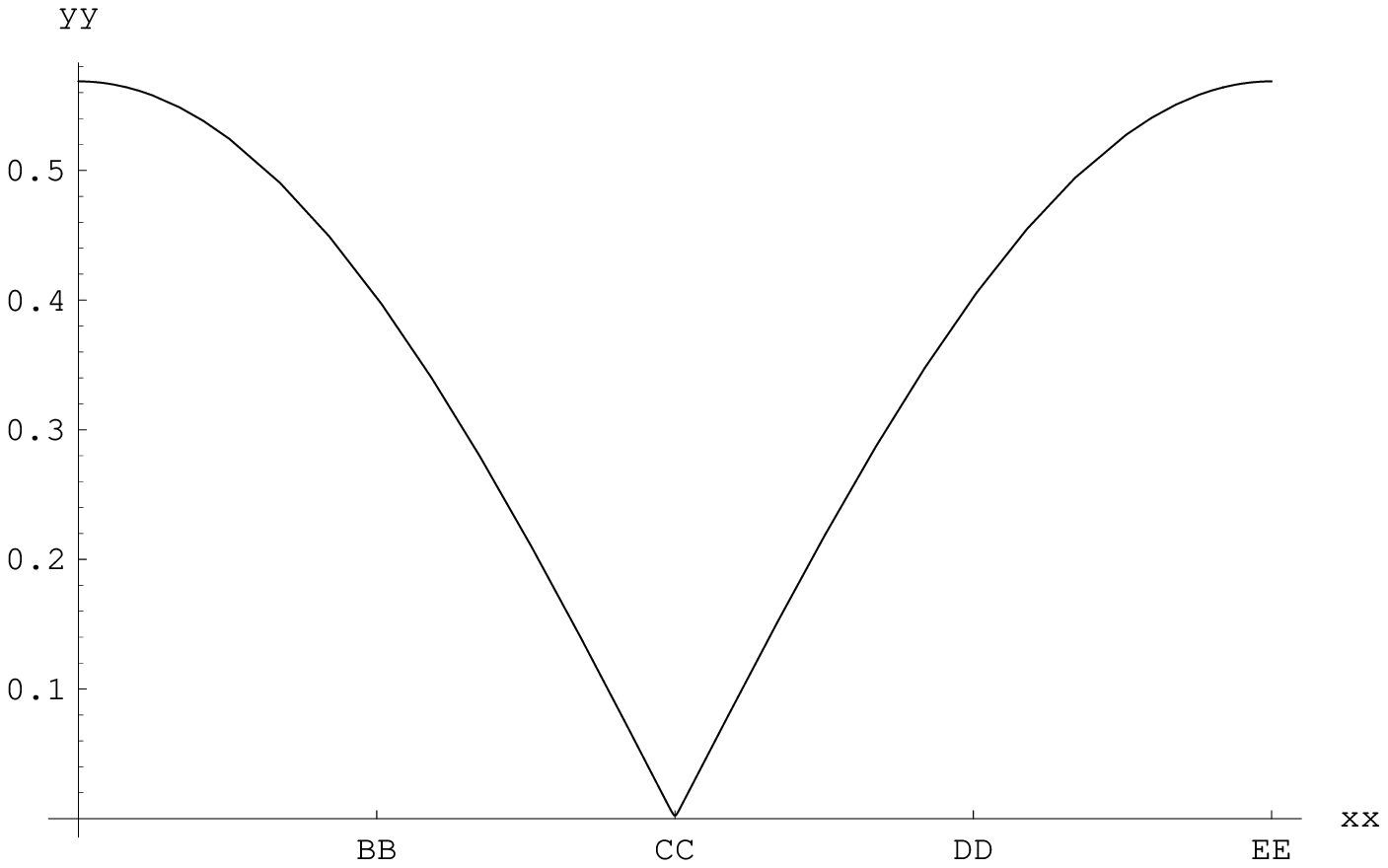}     
  \hspace{0.5in}%
  \fbox{\begin{minipage}{2.5in}     
\psfrag{xx}{\small ${\alpha}$}     
\psfrag{yy}{}     
\psfrag{AA}{$0$}     
\psfrag{BB}{\tiny${23\pi\over 48}$}     
\psfrag{CC}{\tiny${\pi\over 2}$}     
\psfrag{DD}{\tiny${25\pi\over 48}$}     
\psfrag{EE}{\tiny${26\pi\over 48}$}     
\includegraphics[width=2.5in]{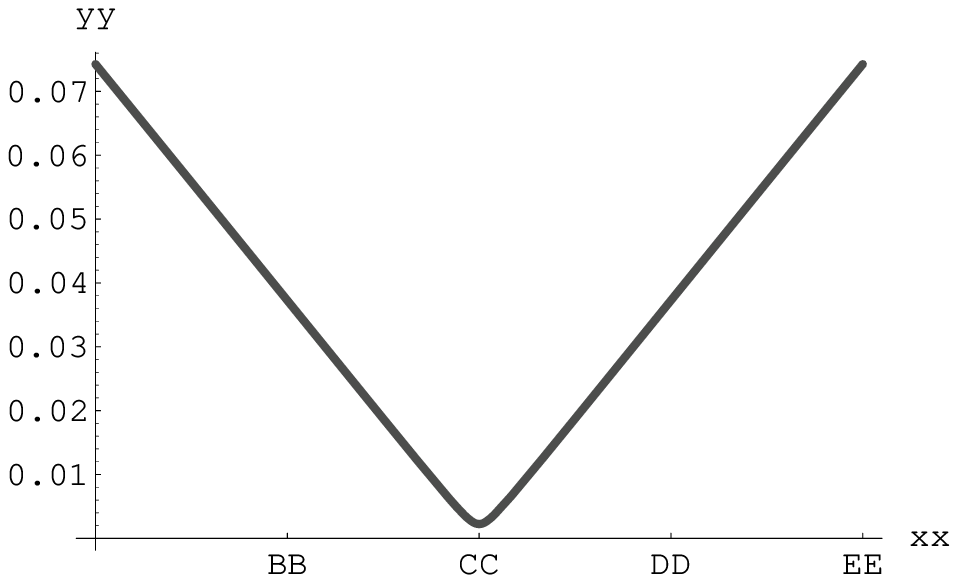}     
 \end{minipage}}          
\caption{\small The normal hierarchy case: the effective mass
normalized as ${\meff/\sqrt{\sun}}$ is plotted against the only CP
phase $\alpha$ (see text). We have used the best fit value for the
solar angle.  In the right panel we zoom the part where there is an
extreme cancellation. }  \end{figure}
 \begin{figure} [t] \begin{center} \psfrag{XX}{\small
   ${\alpha_{\mbox{\tiny {\sc M}}}}$} \psfrag{YY}{} \psfrag{AA}{$0$}
   \psfrag{BB}{\tiny${\pi\over 6}$} \psfrag{CC}{\tiny${2\pi\over 6}$}
   \psfrag{DD}{\tiny${\pi\over 2}$} \psfrag{EE}{\tiny${4\pi\over 6}$}
   \psfrag{FF}{\tiny${5\pi\over 6}$} \psfrag{GG}{\tiny${\pi}$}
   \includegraphics[width=4in]{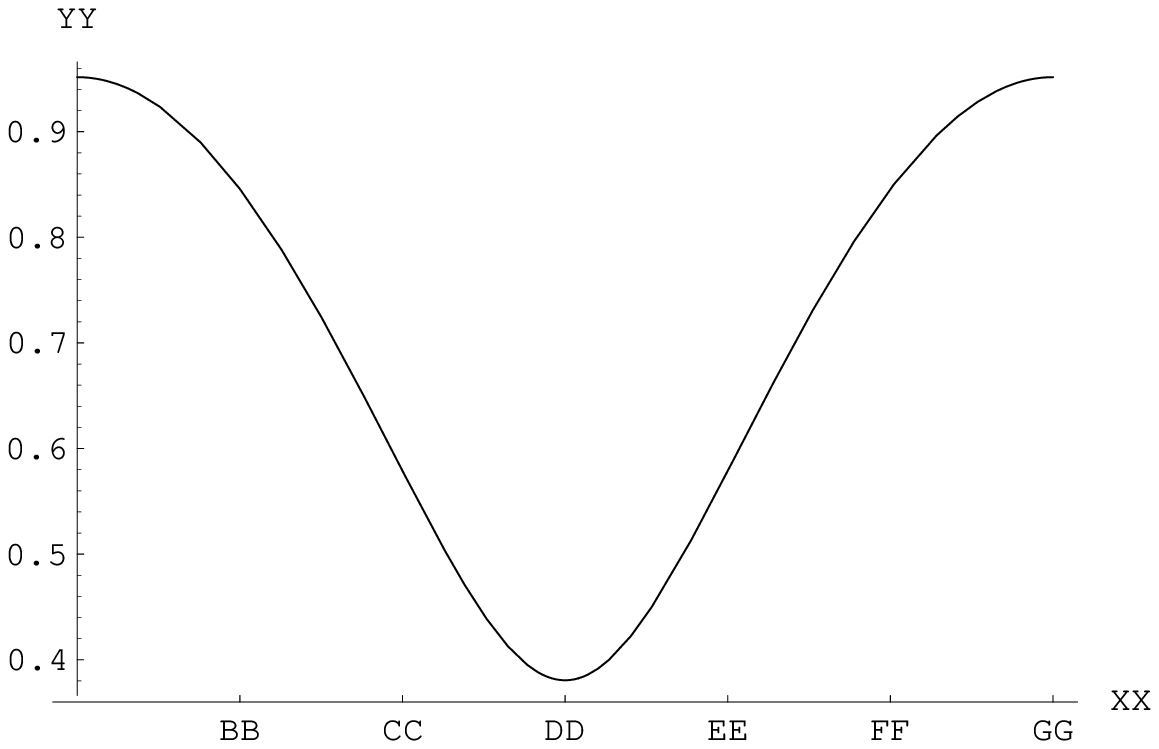} \caption{\small The
   inverted hierarchy case: the effective mass normalized as
   ${\meff/\sqrt{\atm}}$ is plotted against the only CP phase
   $\alpha_{_{M}}$ (see text). We have used the best fit value for the
   solar angle.  }  \end{center} \end{figure}


\begin{thebibliography}{99}     
\bibitem{db_klap} H.V. Klapdor-Kleingrothaus, A. Dietz, H.L. Harney,     
I.V. Krivosheina, Mod. Phys. Lett.  A 16 (2001) 2409.      
     
\bibitem{criticism} C.E. Aalseth {\it et al.}, hep-ex/0202018 and 
F. Feruglio, A. Strumia, F. Vissani, hep-ph/0201291; H.L. Harney, 
hep-ph/0205293. 
     
\bibitem{klap_reply}  H.V. Klapdor-Kleingrothaus, hep-ph/0205228;      
H.V. Klapdor-Kleingrothaus, A. Dietz, I.V. Krivosheina,      
Found. Phys. 32 (2002) 1181.     
     
\bibitem{nemo3} NEMO3 Collaboration, C. Marquet {\em et al.},     
Nucl. Phys. B (Proc. Suppl.) 87 (2000) 298; L. Simard for the NEMO 
collaboration, Nucl. Phys. B (Proc. Suppl.) 110 (2002) 372.     
     
\bibitem{future_db} H.V. Klapdor-Kleingrothaus {\em et al.} (GENIUS),  
J. Phys. G 24 (1998) 483; M. Danilov {\em et al.}, (EXO) Phys. Lett. B  
480 (2000) 12; L. Braeckeleer (for Majorana Collaboration), {\em  
Proceedings of the Carolina Conference on Neutrino Physics}, Columbia  
SC USA, March 2000; E. Ejiri {\em et al.} (MOON), Phys. Rev. Lett. 85  
(2000) 2917; see also, S.R. Elliot, P. Vogel,  
Annu. Rev. Nucl. Part. Sci, 52 (2002).  
     
\bibitem{kps} H.V. Klapdor-Kleingrothaus, H. P\"{a}s, A.Y. Smirnov,      
Phys. Rev D 63 (2001) 073005, and references therein.       
     
\bibitem{petcov02} S. Pascoli, S.T. Petcov, W. Rodejohann,
Phys. Lett. B 558 (2003) 141; {\em ibid.} B 549 (2002) 177. See also,
{\em e.g.}, S. Pascoli, S.T. Petcov, Phys. Lett. B 544 (2002) 239;
S. Pascoli, S.T. Petcov, L. Wolfenstein, Phys. Lett. B 524 (2002) 319;
S.M. Bilenky {\em et al.}, Phys. Rev. D 64 (2001) 053010;
W. Rodejohann, hep-ph/0203214.
       
     
\bibitem{solar_expt} SNO Collaboration, Q.R. Ahmad {\em et al.}, 
Phys. Rev. Lett. 89 (2002) 011301; KamLAND Collaboration, K. Eguchi 
{\em et al.}, Phys. Rev. Lett. 90 (2003) 021802. 
     
\bibitem{solar}     
M.~Maltoni, T.~Schwetz, J.~W.~Valle,     
hep-ph/0212129;     
A. Bandyopadhyay {\em et al.}, hep-ph/0212146;     
J.~N.~Bahcall, M.~C.~Gonzalez-Garcia, C.~Pena-Garay,     
hep-ph/0212147;     
V. Barger and D. Marfatia, Phys.\ Lett.\ B555, 144 (2003);     
P.C. de Holanda, A.Y. Smirnov, hep-ph/0212270; H. Nunokawa {\em et     
al.}, hep-ph/0212202; P. Aliani {\em et al.}, hep-ph/0212212.      
     
\bibitem{atm_expt} Super-Kamiokande Collaboration, Y. Fukuda {\em et     
al.}, Phys. Rev. Lett. 81 (1998) 1562; M. Shiozawa, talk given at     
``Neutrino'02'', Munich, Germany, 2002.       
     
\bibitem{atm}     
N.~Fornengo, M.~Maltoni, R.~T.~Bayo, J.~W.~Valle,     
Phys. Rev. D 65 (2002) 013010.     
     
\bibitem{chooz} CHOOZ collaboration, M. Appolonio {\em et al.}, Phys.     
Lett. B 466 (1999) 415.      
     
\bibitem{pv} Palo Verde experiment, F. Boehm {\em et al.},     
Phys. Rev. Lett. 84 (2000) 3764.     
     
\bibitem{wmap} D.N. Spergel {\em et al.}, astro-ph/0302209.     
     
\bibitem{2df}     
O. Elgaroy et al., Phys. Rev. Lett. 89, 061301 (2002).     
   
\bibitem{gkm} A. de Gouv\^{e}a, B. Kayser, R. Mohapatra, Phys. Rev. D   
67 (2003) 053004.    
     
\bibitem{lsnd} LSND Collaboration, A. Aguilar {\em et al.},      
Phys. Rev. D 64 (2001) 112007.   
   
\bibitem{miniboone} E.D. Zimmerman, hep-ex/0211039, Invited talk at   
the Seventh International Workshop on Tau Lepton Physics (TAU02),   
Santa Cruz, Ca, USA, Sept 2002.   
     
 \bibitem{Green} A. Green, Invited talk at 38th Rencontres de Moriond:   
  Electroweak Interactions and Unified Theories, Les Arcs, France,   
  15-22 Mar 2003.   
   
\bibitem{pmns} B. Pontecorvo, Zh. Eksp. Teor. Fiz. 33 (1957) 549 and   
{\em ibid.} 34 (1958) 247; Z. Maki, M. Nakagawa, S. Sakata,   
Prog. Theor. Phys. 28 (1962) 870.   
     
\bibitem{C-K} L.L. Chau and W.Y. Keung, Phys. Rev. Lett. 53 (1984)      
1802.     
     
\bibitem{bp_rev} S.M. Bilenky, S.T. Petcov, Rev. Mod. Phys. 59 (1987)     
671.     
   
\bibitem{jhf} Y. Itow {\em et al.}, hep-ex/0106019.    
   
\bibitem{numi} D. Ayres {\em et al.}, hep-ex/0210005.    
   
\bibitem{huber} P. Huber, M. Lindner, T. Schwetz, W. Winter,   
hep-ph/0303232.    
     
\bibitem{Fogli} G.L. Fogli {\em et al.}, Phys. Rev. D 66 (2002) 093008.   

\bibitem{forest} R.A. Croft {\em et al.} Astrophys. J. 581 (2002) 20;   
U. Seljak, P. McDonald, A. Makarov, astro-ph/0302571.    
   
    
\bibitem{hannestad} S. Hannestad, astro-ph/0303076.      
    
\bibitem{mainz} C. Weinheimer {\em et al.}, Phys. Lett. B 460 (1999)   
219.    
   
\bibitem{troitsk} V.M. Lobashev {\em et al.}, Phys. Lett. B 460 (1999)   
227.    
    
\bibitem{katrin} KATRIN Collaboration, A. Osipowicz {\em et al.},    
hep-ex/0109033 (letter of intent for next generation Tritium beta    
decay experiment).    
    
\bibitem{am} A. Abada, M. Losada, Nucl. Phys. B 585 (2000) 45.  
 
\bibitem{future} J.L. Vuilleumier, Invited talk at 38th Rencontres de
  Moriond: Electroweak Interactions and Unified Theories, Les Arcs,
  France, 15-22 Mar 2003.

\bibitem{Mohapatra_Pal} `Massive Neutrinos in Physics and
Astrophysics', R.N. Mohapatra and P.B. Pal (World Scientific, 2nd
Edition, 1998).  

\bibitem{postwmap} K. Matsuda, T. Fukuyama, H. Nishiura, 
hep-ph/0302254; K. Cheung, hep-ph/0302265; G. Bhattacharyya, 
H. P\"{a}s, L. Song, T. Weiler, hep-ph/0302191; 
H.V. Klapdor-Kleingrothaus, U. Sarkar, hep-ph/0304032. 
  
     
     
     
     
\end{thebibliography}
\end{document}